\documentclass[twocolumn,showpacs,preprintnumbers,amsmath,amssymb]{revtex4}
\usepackage{tabularx,graphicx}\begin{document}
%\documentstyle[aps]{revtex}
%\documentstyle[preprint,aps]{revtex}
%\begin{document}
\newcommand{\beq}{\begin{equation}}
\newcommand{\eeq}{\end{equation}}
\newcommand{\beqn}{\begin{eqnarray}}
\newcommand{\eeqn}{\end{eqnarray}}
\newcommand{\bmath}{\begin{subequations}}
\newcommand{\emath}{\end{subequations}}
%\draft
\title{Spin currents, relativistic effects and the Darwin interaction in the theory of hole superconductivity}
\author{J. E. Hirsch }
\address{Department of Physics, University of California, San Diego\\
La Jolla, CA 92093-0319}
 
%\date{July 5, 2005} 
\begin{abstract} 
The existence of macroscopic spin currents in the ground state of superconductors is
predicted within the theory of hole superconductivity. Here it is shown that the
electromagnetic Darwin interaction is attractive for spin currents and repulsive for charge currents.
It is also shown that the mere existence of spin currents implies that some electrons are moving
at relativistic speeds in macroscopic superconductors, which in turn implies that the Darwin
interaction plays a fundamental role in stabilizing the superconducting state.

\end{abstract}
%\date{July 5, 2005}
\maketitle 

The theory of hole superconductivity\cite{holeth} predicts that superconductors
expel negative charge from the interior towards the surface\cite{chargeexp}.
As a consequence an excess of negative charge is predicted to exist within a London penetration
depth of the surface and excess positive charge in the interior. A
description of superconductors as 'giant atoms' results\cite{giant},
and an alternative electrodynamic theory\cite{electro} which is different from the
conventional London theory\cite{london} and leads to experimentally testable consequences\cite{tao}.

As a consequence it is predicted that macroscopic spin currents should exist
in the superconducting state\cite{giant,spinc}
 in the absence of applied fields,
as shown schematically in Figure 1. Classically this can be understood from the requirement that
electrons near the surface should move in macroscopic orbits of large angular
momentum so that the centrifugal force prevents them from
'falling in' towards the interior to neutralize the positive charge. Because in the
absence of applied magnetic field there is no charge current in the superconductor, electrons of
opposite spin will orbit in opposite direction. The preferred direction is
determined by spin-orbit coupling, i.e. the electron magnetic moment points
predominantly parallel to its orbital angular momentum. Thus in the absence of applied fields Cooper
pairs $(k\uparrow, -k\downarrow)$ carry no charge current but carry spin 
current.

\begin{figure}
\resizebox{6.5cm}{!}{\includegraphics[width=7cm]{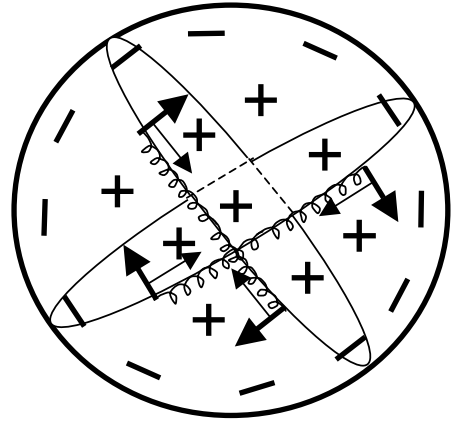}}
\caption{Schematic picture of Cooper pairs in a spherical superconductor giving rise to spin currents. The arrow parallel to the orbit indicates the direction of motion and the arrow perpendicular to the orbit indicates the
direction of the electron magnetic moment  to give minimum spin-orbit energy. Excess negative charge exists within a London penetration depth of the surface, and excess positive charge in the interior.}
\label{figure1}
\end{figure}

Superconductivity is of course a quantum phenomenon.
However the purpose of this paper is to show that  classical
arguments can shed important light into the fundamental physics of superconductors.
 In the sense of Bohr's correspondence principle we
argue that the macroscopic quantum manifestations of superconductivity should also be understandable from
a classical point of view. Here we focus on the Darwin interaction, which is the lowest order
relativistic correction to the electrodynamic Lagrangian of interacting charged particles\cite{darwin1,jackson}.

That the Darwin interaction plays a fundamental role in superconductivity was first proposed in
prescient work by H. Essen in 1995\cite{essen1}, and advocated by him in several papers thereafter\cite{essen2,essen4}.
Essen presents a variety of microscopic and macroscopic arguments in support of his assertion that the Darwin interaction leads to
an attractive interaction between electrons and consequently to Cooper pairing and to superconductivity. 
However, in Essen's treatment the Darwin interaction is found to be attractive
for $parallel$ currents, hence he predicts the existence of $charge$ currents in the ground state of superconductors,
similar to old theories of  Heisenberg\cite{heis} and of Born and Cheng\cite{born}, which is however proven
impossible by 'Bloch's theorem'\cite{bohm}. According to Essen's treatment the interaction
would be repulsive for spin currents, and in contrast to our work there is no charge inhomogeneity in his 
description of superconductors. Furthermore, in Essen's treatment electrons move at non-relativistic speeds and hence
the magnitude of the Darwin energy is small, and certainly cannot account for high $T_c$ superconductivity\cite{essen1}.
Instead we argue here that the interaction is in fact $attractive$ for $spin$ currents and $repulsive$ for
$charge$ currents, and in addition that electrons in spin currents move at relativistic speeds. 
In our
work, no charge currents exist in superconductors in the absence of applied fields, and  
'Bloch's theorem' does not preclude spin currents.\cite{bohm}

The possible role of the Darwin interaction in superconductivity was also considered in the work of  Capelle, Gross
 and coworkers\cite{capelle}. 
However in that work it was not
found that the Darwin term plays an essential role in the attractive interaction leading to pairing, rather relativistic corrections were
considered 'independent of the interaction itself'. The role of the Darwin interaction in superconductivity was also considered
in the work of Lurie and Cremer\cite{lurie}, however its possible role in the pairing interaction was not discussed there either.

There has been confusion in the literature about the sign of the Darwin interaction. In applications to plasma physics,
early work used an approximate form that predicts an attractive interaction for $parallel$ (charge) currents\cite{krisan}.
This was later corrected\cite{trub} and it is now agreed by several workers  that the interaction is in fact attractive for
$antiparallel$ currents\cite{trub,appel, deluca1,deluca2}, while others continue to  argue that the opposite is true\cite{essen3}.

The Darwin electromagnetic interaction Lagrangian between two charges $q_1$ and $q_2$ includes the lowest order relativistic correction and is
given by\cite{jackson}
\beq
L_{int}=\frac{q_1q_2}{r}[-1 + \frac{1}{2c^2}[\vec{v}_1\cdot\vec{v}_2 +    ( \vec{v}_1\cdot \hat{r}) ( \vec{v}_2\cdot \hat{r})   ]]
\eeq
where $\vec{v}_1$, $\vec{v}_2$ are the velocities and $ \hat{r}$ is the unit vector connecting the positions of both charges. The Hamiltonian is
obtained from the Lagrangian through the relation\cite{goldstein}
\bmath
\beq
H=\sum_j v_j p_j -L
\eeq
with
\beq
p_j=\frac{\partial L}{\partial v_j}
\eeq \emath
hence for a time-independent situation where the Hamiltonian gives the constant energy of the system, the interaction energy is
\beq
E_{int}=\frac{q_1q_2}{r}[1 + \frac{1}{2c^2}[\vec{v}_1\cdot\vec{v}_2 +    ( \vec{v}_1\cdot \hat{r}) ( \vec{v}_2\cdot \hat{r})   ]]
\eeq
Note that in the usual situation where $L=T-V$, $H=T+V$, with $T$ and $V$ the kinetic and potential energy, the sign of the second term in
Eq. (1) would have changed in going from Eq. (1) to Eq. (3). This is not the case here because the potential is velocity-dependent.

Inspection of Eq. (3) reveals that the interaction energy between two charges is lowered when their velocities have opposite sign, and 
that the magnitude of the second term becomes important compared to the ordinary Coulomb repulsion for speeds approaching the
speed of light. Both of these facts are key to  the role of the Darwin interaction in superconductivity within our theory.

The confusion about the sign of the Darwin interaction arises because expression (3) appears to contradict the well-known fact
that in magnetostatics parallel charge currents attract. Hence it is argued  that Eq. (3) is 'misleading'\cite{essen3} and instead Eq. (1) should be
interpreted, using $L=T-V$, as representing a potential energy that is lower for parallel currents. In fact, as discussed by
Schwinger\cite{schwinger} the correct expression for the interaction energy in terms of charge and current density for a set of
charges $q_i$ with velocities $\vec{v}_i$ is
\beqn
E&=&\frac{1}{2} \int d^3r d^3 r' \frac{\rho(\vec{r},t)\rho(\vec{r}',t)}{|\vec{r}-\vec{r}'|}+  \nonumber   \\ 
&+& \frac{1}{2c^2}  \int d^3r d^3 r' \frac{\vec{j}(\vec{r},t)\vec{j}(\vec{r}',t)}{|\vec{r}-\vec{r}'|}+ \nonumber   \\
&+&\frac{1}{4c^2}   \int d^3r d^3 r' \frac{\partial}{\partial t} \rho(\vec{r},t)   \frac{\partial}{\partial t} \rho(\vec{r}',t) |\vec{r}-\vec{r}'|
\eeqn
with
\bmath
\beq
\rho(\vec{r},t)=\sum_iq_i\delta(\vec{r}-\vec{r}_i(t))
\eeq
\beq
\vec{j}(\vec{r},t)=\sum_iq_i \vec{v}_i \delta(\vec{r}-\vec{r}_i(t))
\eeq
\emath
and using the continuity equation
\beq
\frac{\partial \rho}{\partial t}=-\vec{\nabla} \cdot\vec{j}
\eeq
Eq. (3) results for each pair of particles. The second term in Eq. (4) also appears to contradict the expectation that the energy
should be lower for parallel currents, however when combining the second and third terms in Eq. (4) Schwinger shows that they reduce
to
\beq
E_{mag}=-\frac{1}{2c^2}  \int d^3r d^3 r' \frac{\vec{j}(\vec{r})\vec{j}(\vec{r}')}{|\vec{r}-\vec{r}'|}
\eeq
predicting lower energy for parallel currents. However in obtaining Eq. (7) from Eq. (6) a boundary term, that is indeed irrelevant for
long parallel wires, is dropped. For our case, Eq. (7) is not applicable.

The following physical argument may help clarify the situation. The interaction between parallel wires with currents of the same sign is
attractive because the magnetic field generated by the wires has opposite sign in the region between the wires. The magnetic field is strongest
close to the wires, and because it contributes to the magnetic energy through
\beq
U_{mag}=\int d^3r \frac{B^2(\vec{r})}{8\pi}
\eeq
it will lower the energy to bring the wires closer together to cancel the magnetic fields in the region close to the surface of each wire facing the other wire.
For the same reason, the force between parallel wires with antiparallel currents is repulsive.
However, consider two identical current loops made of wires of finite cross section so that the planes of the loops are parallel and  their centers lie on a common
 axis perpendicular to the planes of the loops, 
and allow   to change their relative distance $d$ including the regime where the loops overlap. For parallel currents, the magnetic energy for $d=0$ is twice as large
as for $d=\infty$ (four times as large as the energy of a single loop, because the current is twice as large), while for antiparallel currents the magnetic energy when $d=0$ is zero
because the currents have cancelled out. Consequently the interaction in the case of antiparallel currents has to be attractive in some regime (since the energy gets lowered
from finite to zero as $d$ goes from $\infty$ to $0$) and for the case of parallel currents it has to be repulsive in some regime, to increase the energy. Clearly that regime is when the loops are
close and the currents overlap.   We conclude that for charged particles circulating
in the same region of space indeed the energy will be lowered when the particles circulate in opposite directions and   no net charge current exists so that no magnetic field
is generated.

For two electrons of opposite spin orbiting in opposite direction in a classical orbit of radius R the energy Eq. (3) is
\beq
E_{int}(\theta)=\frac{e^2}{r}[1-\frac{v^2}{c^2} \frac{1+3 cos\theta}{4}]
\eeq
where $\theta$ is the angle between the position vectors of both charges which changes with time according to $\theta = 2\omega t$,
$\omega=v/R$, and $r$ is the instantaneous distance between the charges, $r=\sqrt{2}R\sqrt{1-cos\theta}$. Note that
when $\theta \rightarrow 0$ the second term in Eq. (9) exactly cancels the first term, the ordinary Coulomb repulsion,
as $v\rightarrow c$. For any angle $0\leq \theta \leq 109.5^o  $ the velocity-dependent term is negative and reduces the
Coulomb repulsion. Small values of $\theta$ correspond of course to small $r$ which is when the ordinary Coulomb repulsion is
largest.

In the model of hole superconductivity a uniform positive charge distribution exists in the interior of superconductors giving rise
to a radial electric field $E=cr$, with $c$ of order $10^6 V/cm^2$\cite{chargeexp}. The angular velocity of electrons is then
$\omega=(ec/m_e)^{1/2}$ with $m_e$ the electron mass, and the second term in Eq. (9) will become very important for
macroscopic samples.

In a quantum-mechanical treatment, assuming a spherical geometry the Cooper pair can be denoted by
\beq
c_{nlm\uparrow}^\dagger c_{nl-m\downarrow}^\dagger
\eeq
with $n$, $l$, $m$ radial, orbital and azimuthal quantum numbers. It represents electrons orbiting in opposite directions as in the
classical example discussed above, hence we conclude that the energy of a Cooper pair is lowered by the velocity-dependent
Darwin interaction. There is yet another relativistic  lowering of the energy of Cooper pairs, from the spin-orbit interaction
with the internal electric field $\vec{E}$
\beq
E_{s.o.}=\frac{e}{2 m_e c^2}\vec{S}\cdot (\vec{v}\times\vec{E})
\eeq
with $\vec{S}$ the electron spin. Even though this term goes as $(v/c)$ compared to the Darwin term's 
$(v/c)^2$ dependence, the prefactor is small and at relativistic speeds the energy lowering from this term is negligible
compared to the Darwin term.
 
 The expression Eq. (9) diverges as $\theta\rightarrow 0$, hence cannot be used to estimate the energy. As a simple example, consider
two spherical shells of radius $R$ and charge $q$ rotating in opposite direction with angular velocity $\omega$, corresponding to the
electrons with spin 'up' and 'down' respectively. In a stationary situation
the third term in Eq. (4) vanishes and we can calculate the Darwin energy from just the second term in Eq. (4)
\bmath \beq
E_{Darwin}= \frac{1}{2c^2}  \int d^3r d^3 r' \frac{\vec{j}_\uparrow(\vec{r})\vec{j_\downarrow}(\vec{r}')}{|\vec{r}-\vec{r}'|}
\eeq
with
\beq
\vec{j_\sigma}(\vec{r})=\frac{q}{4\pi R^2}\sigma \vec{\omega}\times \vec{r}
\eeq
\emath
and $\sigma=\pm1$, and obtain
\beq
E_{Darwin}=-\frac{q^2}{9}\frac{\omega^2 R}{c^2}=-\frac{N^2 e^2 \omega^2 R}{9c^2}
\eeq
for $q=Ne$, with $N$ the number of Cooper pairs. The kinetic energy is (neglecting relativistic corrections)
\beq
E_{kin}=\frac{2}{3} N m_e \omega^2 R^2
\eeq
so that for large $N$ the Darwin energy lowering is greater than the kinetic energy associated with the spin current.
The relevant parameter (ratio of Eqs. (13) and (14))  is $Ne^2/m_e c^2 R$ as already found by   Essen\cite{essen4}. 
For relativistic speeds the energy lowering Eq. (13) is of the same order of magnitude as the Coulomb energy 
cost of the inhomogeneous charge distribution\cite{chargeexp}.
 
 Note also that the rigid rotation assumed in Eq. (12b) is far from optimal to benefit from the Darwin energy lowering, since only the
 electrons in the equatorial plane are moving at the maximum speed $v=\omega R$. Other patterns of motion where 
 electrons move along various great circles of different orientations as depicted in Fig. (1) will yield larger energy lowering than
 Eq. (13). In ref. \cite{deluca2}, a numerical simulation of electrons on the surface of a sphere was performed and it was found that electrons
 spontaneously develop complicated motion patterns that are indeed characterized by neighboring electrons having velocities
 in opposite directions.
 
 In the theory of hole superconductivity, a negative charge density $\rho_-$ exists within a London penetration depth of the
 surface, and a positive charge density $\rho_0$ in the interior. The charge density obeys the differential equation
 \bmath
 \beq
 \rho(\vec{r})=\rho_0+\lambda_L^2\nabla^2\rho(\vec{r})
 \eeq
 with $\lambda_L$ the London penetration depth, and the electric field is determined by the equation
 \beq
 \nabla^2(\vec{E}-\vec{E}_0)=\frac{1}{\lambda_L^2} (\vec{E}-\vec{E}_0)
 \eeq
 where $\vec{E}_0$ is the electric field originating in a uniform charge density $\rho_0$. We consider a spherical
 geometry for simplicity, where $\vec{E}_0$ is given by
 \beq
 \vec{E}_0=\frac{q}{R^3}\vec{r}
 \eeq
 \emath
 and $q=4\pi R^3\rho_0/3$ is the charge expelled from the interior of the sphere of radius $R$ to the surface shell. 
 The Darwin interaction term becomes important compared to the ordinary Coulomb repulsion when the speed 
 approaches the speed of light. We now show that in fact the predicted spin currents\cite{spinc} involve such speeds for samples
 of dimensions much larger than the London penetration depth.
 
 Consider a sphere of radius $R>>\lambda_L$, and electrons associated with the spin current near the
 surface rotating with azimuthal velocity $v_0$, in opposite directions for spin $\uparrow$ and $\downarrow$.
 Mechanical equilibrium requires that
 \beq
 \frac{m_e v_0^2}{r}=eE(r)
 \eeq
 for electrons at radius $r$. We will use nonrelativistic expressions for simplicity since it is sufficient to illustrate the
 point. When a magnetic field $\vec{B}$ is applied, the speed of electrons within a London penetration depth of the
 surface is modified to screen the magnetic field, denote the change in speed by $v_\phi$. To maintain mechanical equilibrium to first order it is seen from variation of Eq. (16) that the condition
 \beq
 v_\phi=-\frac{e}{m_ec} \frac{Br}{2}
 \eeq
 is required. Now from integrating the London equation
 \beq
 \vec{\nabla}\times\vec{J}=-\frac{c}{4\pi\lambda_L^2}\vec{B}
 \eeq
 we can write the charge current $\vec{J}$ as
 \bmath
 \beq
J\sim \frac{c}{4\pi\lambda_L}B
 \eeq
 and defining the current density as
 \beq
 \vec{J}=\rho\vec{v}_\phi
 \eeq
 \emath
 we conclude that the charge density $\rho$ that contributes to the current is given by
 \beq
 \rho=\frac{m_e c^2}{2\pi\lambda_L  er}
 \eeq
 where $r\sim R$ to within a London penetration depth correction. Using for the London penetration depth
 \beq
 \frac{1}{\lambda_L^2}=\frac{4\pi n_s e^2}{m_e c^2}
 \eeq
 with $n_s$ the density of electrons in the condensate, Eq. (20) yields
 \beq
 \rho=2e n_s \frac{\lambda_L}{r}
 \eeq
 which differs in a fundamental way from the usual assumption in London theory. Namely, it says that only a small
 fraction $\sim (\lambda_L/r)$ of all electrons contribute to the transport current, with a speed that is larger by
 a factor $r/\lambda_L$ from the speed assumed in conventional London theory.
  
 The maximum value that the electric field attains is $E_{max}\sim q/R^2\sim 4\pi \lambda_L \rho_-$ within distance
 $\lambda_L$ of the surface\cite{chargeexp}. Here, $q$ is the expelled charge and $\rho_-$ is the excess negative
 charge density near the surface. This electric field determines the maximum speed of the spin current 
 $v_0$ (Eq. (16) for $r \sim R$). Under the reasonable assumption that the transport charge density $\rho$ given by eq. (20) is
 of the same order of magnitude as the expelled charge density $\rho_-$, one is led from Eqs. (16) and
 (20) to the startling
 conclusion that $v_0 \sim c$ in samples of dimensions much larger than $\lambda_L$\cite{explain}. This then implies that
 the Darwin interaction is all-important in the energy balance.

 In addition, the condition $\rho=\rho_-$ leads through Eq. (20)  to the remarkable conclusion that
 \beq
\frac{qe}{R}=2m_e c^2
\eeq
in other words, that the potential energy of electrons within $\lambda_L$ of the surface where the electric field is
maximum is right at the threshold where pair production becomes energetically possible!\cite{diracatom,bcs}. Possible experimental
consequences of this finding will be discussed elsewhere.

Finally we may ask, how many Cooper pairs participate in the charge and spin currents near the surface? Denoting by $N$ the number,
we have $Ne=2\pi R^2 \lambda_L \rho$, and using Eq. (20) for $\rho$
\beq
N=\frac{m_e c^2 R}{e^2}=\frac{R}{r_e}
\eeq
with $r_e$ the classical electron radius. This is precisely the condition derived by Essen\cite{essen4} for the Darwin interaction
to be important. Furthermore, as already showed by Essen\cite{essen4}, it is the condition required to explain
the magnitude of magnetic fields in rotating superconductors (London field)  under the assumptions that only a fraction
$n_s \lambda_L/R$ of electron density (Eq. (22)) near the surface contribute to the current and that these electrons are 
completely unaffected by the body' s rotation (these assumptions were   not made
explicit in Essen's work). In contrast, in the conventional London treatment\cite{london} all the $n_s$ electron density near the 
surface contributes to the London field through a velocity 'lag' $\Delta v \sim \omega \lambda_L$ for each carrier, with $\omega$ the
angular velocity of body rotation.

In summary, we have given a series of arguments that lead to the conclusion that relativistic effects play a
fundamental role in superconductors of dimensions larger than the London penetration depth within the theory
of hole superconductivity. This was already foreshadowed by the prediction\cite{chargeexp} that macroscopic
electric fields of magnitude of order $10^6 V/cm$ exist in the interior of superconductors, which implies
relativistic effects in samples of dimension of order $cm$ (recall that the electron rest energy is $\sim 0.5 \times 10^6 eV$).
Separately, the mere  $existence$ of charge expulsion and spin currents and the condition
of mechanical equilibrium was shown here to imply that electrons near the surface move
at relativistic speeds, which in turn implies that the Darwin electromagnetic interaction has to play an important
role. We have argued that the Darwin interaction is attractive for spin currents and repulsive for charge
currents (in contradicion to previous work), and hence conclude that it   plays a fundamental role in the
stabilization of the superconducting state in the theory of hole superconductivity where spin currents are predicted to
exist. The fact that the Darwin energy lowering is a $kinetic$ energy lowering mechanism was also foreshadowed
by the lattice formulation of the model of hole superconductivity\cite{holemodel}.
Further development of these concepts   will be given in future work.


\begin{references}
 \bibitem{holeth} See www.physics.ucsd.edu/$\sim$jorge/hole.html for a    list of references.
\bibitem{chargeexp} J.E. Hirsch, Phys.Lett. A{\bf 281}, 44 (2001); Phys.Rev. B{\bf 68}, 184502 (2003).
\bibitem{giant} J.E. Hirsch, Phys.Lett. A {\bf 309}, 457 (2003).
\bibitem{electro}  J.E. Hirsch,  Phys.Rev. B{\bf 69}, 214515  (2004).
\bibitem{london} F. London, 'Superfluids', Dover, New York, 1961.
 \bibitem{tao} J.E. Hirsch, Phys. Rev. Lett. {\bf 92} , 016402 (2004); Phys. Rev. Lett. {\bf 94} , 187001 (2005).
\bibitem{spinc}  J.E. Hirsch,  Phys.Rev. B{\bf 71}, 184521  (2005).
\bibitem{darwin1}  C.G. Darwin, Phil. Mag. {\bf 39}, 537 (1920).
\bibitem{jackson}  J.D. Jackson, Classical Electrodynamics, Third Edition, Wiley, New 
York, 1999, p. 596.
\bibitem{essen1} H. Essen, Physica Scripta {\bf 52}, 388 (1995).
\bibitem{essen2} H. Essen, Phys.Rev. E{\bf 53}, 5228  (1996); Phys.Rev. E{\bf 56}, 5858  (1997);  
J. Phys. A{\bf 32}, 2297 (1999).
\bibitem{essen4} H. Essen,    cond-mat/0002096 (2000).
\bibitem{heis} W. Heisenberg, Zeits. f. Naturforschung 32, 65 (1948).
\bibitem{born} M. Born and K.C. Cheng, Nature 161, 1017 (1948).
\bibitem{bohm} D. Bohm, Phys. Rev. 75, 502 (1949).
\bibitem{capelle} K. Capelle,  Phys.Rev. B{\bf 63}, 052503  (2001);  K. Capelle and E.K.U. Gross,   Phys.Rev. B{\bf 59}, 7155  (1999) and references therein.
\bibitem{lurie} D. Lurie and S. Crever, Prog. Theor. Phys. {\bf44}, 300 (70); S. Crever and D. Lurie, J. Low Temp. Phys.
{\bf 3}, 439 (1970).
\bibitem{krisan}  J.E. Krisan and P. Havas,  Phys.Rev. {\bf 128}, 2916  (1962).
\bibitem{trub}  B.A. Trubnikov and V.V. Kosachev, Sov. Phys. JETP {\bf 27}, 501 (1968).
\bibitem{appel} W. Appel and A. Alastuey, Physica A {\bf 252}, 238 (1998).
\bibitem{deluca1} V. Mehra and J. De Luca, , Phys.Rev. E{\bf 61}, 1199  (2000).
\bibitem{deluca2} J. De Luca, S.B. Rodrigues and Y. Levin, cond-mat/050661 (2005).
\bibitem{essen3}  H. Essen and A. Nordmark, Phys.Rev. E{\bf 69}, 036404  (2004).
\bibitem{goldstein} H. Goldstein, C. P. Poole, and J. L. Safko `Classical Mechanics', Third Edition,  Addison Wesley, Boston, 2002.
\bibitem{schwinger} J. Schwinger, L.L DeRaad, K.A. Milton and W.Y. Tsai, `Classical Electrodynamics', Perseus Books, Reading, 1998,
Chpt. 33.
\bibitem{explain} More precisely, using the correct relativistic expression in place of Eq. (16), $v_0=0.91c$.
\bibitem{diracatom} S.S. Gershtein and Y.B. Zeldovich, Sov. Phys. JETP {\bf 30}, 358 (1970).
\bibitem{bcs} The natural connection of this finding to the form of the BCS wave function has not escaped our attention.
\bibitem{holemodel} J.E. Hirsch,  Physica C {\bf 199}, 305  (1992); J.E. Hirsch and F. Marsiglio,
 Phys.Rev. B{\bf 62}, 15131  (2000).

\end{references}
\end{document}